%% file: Main.tex
\documentclass[conference]{IEEEtran}
\IEEEoverridecommandlockouts
\usepackage{amsmath,amssymb,amsfonts}
\usepackage{hyperref}
\hypersetup{
    colorlinks=true,
    linkcolor=blue,
    filecolor=magenta,      
    urlcolor=blue,
    pdftitle={Nested Inheritance Dynamics},
    pdfpagemode=FullScreen,
    }
\usepackage{algorithm}
\usepackage{algpseudocode}
\usepackage{graphicx}
\usepackage{textcomp}
\usepackage{xcolor}
\usepackage{comment}
\usepackage{array}
\usepackage{booktabs}
\usepackage{lipsum}
\usepackage{float}
\usepackage{cite}
\usepackage{enumitem}
\newcommand{\eq}[1]{\begin{equation} #1 \end{equation}}
\newcommand{\bx}{\textbf{x}}

\newcommand{\bX}{\textbf{X}}
\newcommand{\bY}{\textbf{Y}}

\def\BibTeX{{\rm B\kern-.05em{\sc i\kern-.025em b}\kern-.08em
    T\kern-.1667em\lower.7ex\hbox{E}\kern-.125emX}}
\begin{document}

\title{Nested Inheritance Dynamics


}

\author{\IEEEauthorblockN{ Bahman Moraffah}
\IEEEauthorblockA{\textit{School of Electrical, Computer, and Energy Engineering} \\
\textit{Arizona State University}, Tempe AZ, USA\\
Bahman.Moraffah@asu.edu}
}
\maketitle

\begin{abstract}
The idea of the inheritance of biological processes, such as the developmental process or the life cycle of an organism, has been discussed in the biology literature, but formal mathematical descriptions and plausible data analysis frameworks are lacking. We introduce an extension of the nested Dirichlet Process (nDP) to a multiscale model to aid in understanding the mechanisms by which biological processes are inherited, remain stable, and are modified across generations. To address these issues, we introduce Nested Inheritance Dynamics Algorithm (NIDA). At its primary level, NIDA encompasses all processes unfolding within an individual organism's lifespan. The secondary level delineates the dynamics through which these processes evolve or remain stable over time. This framework allows for the specification of a physical system model at either scale, thus promoting seamless integration with established models of development and heredity. 
\end{abstract}


\input{Intro}

\input{Background}

\input{Proposed}
\input{Experiment}

\input{Related_Work}

\section{Conclusion}
This paper introduces NIDA, a novel Bayesian formulation of inheritance that leverages the flexibility and hierarchical nature of the nDP. NIDA offers a structured yet adaptable approach to understanding multi-scale biological phenomena. By employing the nDP, NIDA provides a robust statistical framework for unraveling the intricate processes underlying biological inheritance. NIDA facilitates a nuanced analysis of genetic trait and developmental pattern transmission across generations, accommodating the inherent variability and complexity of biological systems.

\bibliographystyle{unsrt} 
\bibliography{refs} 

\begin{algorithm}[t!]
\caption{NIDA: Nested Inheritance Dynamics Algorithm }
\label{alg:NIDA}
\begin{algorithmic}[1]
\State \textbf{Input:} Prior distributions for parameters
\State \textbf{Output:} Updated parameters

\State \textbf{Initialization:} Initialize parameters based on prior distributions.

\For{each generation and time step}
    \State Update parameters $x_{t,k,d,n}$, $\theta_{t,k,d,n}$ and $\phi_{t,d}$ using MCMC or variational inference based on observed data likelihood (e.g., Hamiltonian Monte Carlo)
    \State Integrate out or sum over latent variables and previous states according to state transition probabilities and observation models.
\EndFor

\State \textbf{Indicators for MCMC Sampling:}
    \State Utilize indicators for each cluster assignment within the nDP to track parameters influenced by hierarchical structure.
\end{algorithmic}
\end{algorithm}

\input{Appendix}

\end{document}

%% file: Intro.tex
\section{Introduction}

Understanding the inheritance and evolution of biological processes across generations is a profound challenge that intersects the fields of genetics, development biology, and statistical modeling. Traditional models often struggle to capture the complexity of multi-generational changes due to their simplistic assumptions about independence and stationarity, \cite{visscher2008heritability}. This paper introduces the Nested Inheritance Dynamics Algorithm (NIDA), a novel  algorithm utilizing nested Dirichlet processes (nDPs) to elucidate the dynamics of biological inheritance that evolve over time and across generations. By leveraging the flexibility of Bayesian nonparametrics through the Nested Dirichlet Process and the structured temporal modeling capabilities of Multiscale Hidden Markov Models, this model provides a robust framework for understanding complex genetic inheritance patterns and developmental trajectories in a multilayered temporal context.

The complexity of genetic data, marked by high dimensionality and the interplay between genes and environmental factors, necessitate advanced statistical approaches that go beyond conventional models. Lynch and Walsh highlight the complexities in modeling genetic variation which are compounded when considering gene-environment interactions and non-Mendelian inheritance patterns, \cite{lynch1998genetics, falconer1983quantitative}. To address these, NIDA leverages Bayesian nonparametrics, particularly nested Dirichlet Processes, \cite{moraffah2024BNP, rodriguez2008nested, hjort2010bayesian}, providing a flexible and robust framework for understanding complex genetic and phenotypic patterns within a multilayered temporal context.

The NIDA specifically addresses the challenges posed by the hierarchical and evolving nature of biological systems, where changes occur both within the lifespan of an organism's lifespan and across lineages. This dual-scale approach allows the model to capture dependencies not only within a single organism’s developmental process but also across successive generations, thus providing insights into both microevolutionary changes and macroevolutionary patterns. This methodological innovation offers a significant advancement over traditional approaches by integrating the structured temporal modeling capabilities of multiscale models, \cite{deisboeck2011multiscale, shoemaker1999bayesian}, with the probabilistic rigor of Bayesian nonparametrics. To our knowledge, this is the first study to define biological inheritance using nonparametric Bayesian methods to explore the dynamics of inheritance. This paper presents three primary contributions:
\begin{itemize}
    \item Introducing a novel probabilistic interpretation of inheritance and integration of Nested Dirichlet Processes, enabling the analysis of complex biological data across multiple scales of temporal resolution.
    \item Providing a new methodological framework for dissecting the mechanisms of genetic inheritance and evolutionary stability.
    \item Offering a highly flexible tool for the analysis of time-series genetic data, capable of accommodating various types of biological data and experimental designs.
\end{itemize}

The paper is structured as follows: Section \ref{sec:back} offers a comprehensive background on the dynamics of inheritance and the nested Dirichlet process. In Section \ref{sec:proposed}, we introduce the mathematical modeling of the inheritance and then propose the Nested Inheritance Dynamics Algorithm (NIDA) algorithm to deepen the understanding of biological inheritance dynamics. We present the model training tools and approximate the underlying distribution of inheritance to predict future inheritance. Through experiments, We demonstrate the efficacy of our nonparametric model, Section \ref{sec:exp}.

%% file: Background.tex
\section{Background}
\label{sec:back}
\subsection{Dynamics of Biological Inheritance}
Analyzing and modeling biological inheritance pose significant mathematical challenges due to the inherent complexity and variability of genetic processes. Biological inheritance encompasses the transmission of genetic material across generations, influencing traits that are fundamental to an organism’s development and survival. The key mathematical challenges in this area stem from the need to accurately model genetic variation, gene-environment interactions, and the transmission of traits that may not follow simple Mendelian patterns of inheritance.

One of the primary mathematical challenges is dealing with the high dimensionality of genetic data. Each organism’s genome contains thousands to billions of base pairs, and modern genomic studies often involve sequencing and comparing large numbers of genomes. Handling this massive amount of data requires robust statistical methods and computational algorithms to identify patterns of inheritance, associate genetic variants with traits, and predict outcomes. Techniques such as principal component analysis (PCA) and machine learning models like random forests and neural networks are employed to reduce dimensionality and improve the manageability and interpretability of genetic data.

Another challenge is the modeling of polygenic traits as traits are influenced by many genes, each contributing a small effect. Unlike single-gene disorders, polygenic traits require the use of quantitative genetic models that can capture the combined influence of multiple genes. Statistical approaches like genome-wide association studies (GWAS) are used to link specific genetic variants to traits, but they must contend with issues like statistical power and false discovery rates due to multiple testing. Moreover, the interaction between genes (epistasis) and between genes and environmental factors (gene-environment interactions) adds another layer of complexity, requiring sophisticated models that can integrate various data types and sources.

The non-linear dynamics of gene expression over time and across different biological conditions also present substantial challenges. Mathematical models must be flexible enough to represent the dynamic processes governing gene expression and its impact on phenotypic traits. This is particularly relevant in the study of developmental processes and life cycles, where gene expression can significantly vary at different stages or in response to environmental changes. Stochastic modeling and dynamic systems theory are often applied in these cases, using differential equations to model the regulatory networks that control gene expression.

The mathematical modeling of biological inheritance has, however, not been vastly addressed. These methods often focus on advancements in statistical genetics, the development of new computational tools for data analysis, and the theoretical underpinnings of genetic inheritance models. 
In this paper, we provide a framework for modeling the inheritance patterns across multiple scales, allowing us to account for variability within and between individual organisms over time. In particular, we employ a nested Dirichlet process prior, and multiscale extension of Hidden Markov Models (HMMs) to model the latent structures in genetic data, capturing the transitions between different states of a genetic trait that are not directly observable. We then utilize a scalable Markov chain Monte Carlo (MCMC) technique to sample from the posterior and estimate the inheritance. This approach facilitates the integration of diverse data sources and improves the robustness of genetic predictions.

\subsection{Mathematical Background}

The nested Dirichlet Process (nDP) and the nested Chinese Restaurant Process (nCRP) are advanced Bayesian nonparametric models used primarily for hierarchical clustering where not only the data within each group is clustered, but the groups themselves can also be clustered in a hierarchical fashion, \cite{moraffah2024BNP, rodriguez2008nested, griffiths2003hierarchical}. These models extend the flexibility and applicability of the classical Dirichlet Process (DP) and the Chinese Restaurant Process (CRP) by introducing additional layers of hierarchy and allowing for more complex dependencies between data points.

\noindent  \textbf{\textit{Nested Dirichlet Process (nDP)}}

The nDP is a hierarchical extension of the Dirichlet Process (DP). In the classical DP setup, a DP is parameterized by a base distribution \( Q \) and a concentration parameter \( \gamma \), which controls the degree of clustering (or the dispersion of clusters). The DP can be visualized as generating distributions where observations are clustered around a set of latent parameters. The nDP takes this concept further by allowing these latent parameters themselves to be drawn from another DP, effectively creating a hierarchy of DPs. The nDP model can be defined through the following steps.

1. Global Base Distribution, \( G_r \), represents a global-level Dirichlet process that acts as the base distribution for other DPs in the hierarchy and is given by
  \eq{G_r \mid \gamma, Q \sim \text{DP}(\gamma, Q),}
where \( Q \) is a base distribution that controls the overall characteristics of the parameters and \( \gamma \) is the concentration parameter that influences the number of distinct clusters at this global level. The Stick-breaking Construction of \( G_r \) is then given by
   \eq{
   G_r = \sum_{l=1}^{\infty} w_{lr} \delta_{\phi_{lr}^*}, \quad \phi_{lr}^* \sim Q, \quad (w_{lr})_{l=1}^\infty \sim \text{GEM}(\gamma).
   }
   The global mixture weights \( w_{lr} \) follow a GEM distribution {\footnote{The GEM (Griffiths-Engen-McCloskey) distribution is a probability distribution defined on the unit interval [0, 1]. Weights \( \{w_{lr}\}_l \sim \text{GEM}(\alpha)\) as \(w_{lr} = v_l \prod_{i=1}^{l-1} (1 - v_i)\)
where \( v_i \sim \text{Beta}(1, \alpha)\) are independent samples and \( \alpha \) is a parameter of the Dirichlet process, \cite{moraffah2024BNP}.}}, which is typical for a stick-breaking process of the Dirichlet Process. And, \( \delta_{\phi_{lr}^*} \) represents a Dirac delta function centered at \( \phi_{lr}^* \), which is drawn from base distribution, \( Q \).

2. Group-specific Distributions, \( \{G_j\}_j \) is modeled as DP with a concentration parameter \( \alpha \) and the global DP \( G_r \) as its base distribution,
  
   \eq{G_j \mid \alpha, G_r \sim \text{DP}(\alpha, G_r).}
This setup allows the sharing of clusters among different groups, facilitating the borrowing of statistical strength across them.

3. Individual-specific parameters \( \theta_{ji} \) are drawn from the group-specific distribution \( G_j \). These parameters dictate the clustering of observations within the group,
  
  \eq{ \theta_{ji} \mid G_j \sim G_j.}

4. Observations \( x_{ji} \) within each group are generated from a distribution \( F \) parameterized by \( \theta_{ji} \), which are drawn from the corresponding group-specific DP as follows
\eq{ x_{ji} \mid \theta_{ji} \sim F(\theta_{ji}).}

\noindent \textbf{\textit{Nested Chinese Restaurant Process (nCRP)}}

 The nCRP is a process-based interpretation of the nDP, extending the analogy of the Chinese Restaurant Process used to describe the behavior of the classical Dirichlet Process \cite{moraffah2024BNP}. In CRP, customers (data points) sit at tables (clusters) in a restaurant (the dataset), and the arrangement of customers at tables follows a specific probabilistic rule based on the number of customers already seated at each table.
In the nCRP, the restaurant analogy is expanded to a franchise of restaurants, each with its own local seating arrangement but also connected by a global menu shared across the franchise:

1. Global Restaurant:
   Customers first choose a franchise location based on the popularity and the characteristics of each franchise, governed by a global DP.

2. Local Restaurants:
   Within each franchise location, customers choose tables according to a local CRP. The choice of table not only depends on the number of customers already seated but also on the characteristics of the tables that are influenced by the global menu.

This nested structure allows for a complex hierarchical clustering where data points are clustered within their local groups, and these groups themselves are clustered at a higher level, all within a probabilistic framework that facilitates the sharing of statistical strengths across different parts of the data.

%% file: Proposed.tex
\section{Proposed Method}
\label{sec:proposed}
Biological processes and their evolutionary pattern exhibit complex hierarchical structures that are challenging to model and analyze. Traditional methods often fail to capture the full spectrum of biological variability and inter-generational dependencies. In this section, we introduce a Bayesian nonparametric approach to address these challenges. The model particularly focuses on the inheritance patterns of gene expression and morphogenesis, which are represented as time-series data within a multiscale system. The objective is to estimate state and transition probabilities, providing insights into both genetic expressions at various developmental stages and their evolution over generations. We discuss the theoretical understanding of this approach and its application in biological inheritance and evolution. In particular, the nDP is applied as a prior over parameters such as the mean and covariance matrix, which are pivotal in modeling complex biological systems.

\noindent\textbf{Notations:} Let \(h_{t,k, (N)}\) denote the set \(\{h_{t,k,1}, \dots, h_{t,k N}\}\). Similarly, \(h_{t, (K), (N)} := \{\{h_{t, k, n}\}_{k=1}^K\}_{n=1}^N\). 

\subsection{Developmental State Modeling: Fine Scale}
\label{sec:fine}

The developmental state of an individual \(d = 1, \dots, D\) is modeled as a multidimensional vector capturing the expression levels of genes at specific developmental time points \(k = 1, \dots, K\) and across generations \(t = 1, \dots T\) and defined as \( \bx_{t, k,d, (N)}\), where \( x_{t, k, d, n} \) represents the expression level of the \(n\)-th gene at the \(k\)-th developmental time point for individual \( d \) in the fixed generation \( t \). These states are captured over \( D \) individuals, \( N \) genes, \( K \) developmental time points, and \( T \) generations.

\textbf{State Transition and Observation Model.}
The process is modeled through a state space model using recursive functions involving the previous states and an innovation term enhancing the model's ability to capture the dynamic biological processes. The state transition and observations for a specific individual \(d\) in fixed generation \(t\), are modeled as 
\begin{flalign}
\begin{split}
       x_{t, k,d, n} &= f(A_{t, k, (N)} x_{t, k-1, d, (N)}, w_{t, k,d, n})\\
 y_{t, k, d, n} &= g(x_{t, k,d, n}, v_{t,k,d,n}),
\end{split}
\end{flalign}
for all  \(n = 1, \dots, N\), \(k =1,\dots, K\), and \(d=1, \dots D\) where \( A_{t, k, (N)} \in \{0,1\}^N\) represents the binary gene transition vector; indicating which genes contribute in transitioning from time point \(k-1\) to time point \(k\), and \( w_{t,k,d, n} \) and \( v_{t,k,d, n} \) are the process and observation noise terms, respectively. Functions \(f\) and \(g\) are functions that determine the passage of genetic information for individuals.

\subsection{Heredity and Evolution: Coarse Scale}
\label{sec:coarse}
To capture the hereditary properties, the model iterates over time and considers the entire developmental trajectory. The model iterates over all individuals' developmental trajectories across all time points \( k \) and generations \( t \) to study the inheritance patterns. The developmental trajectory at generation \( t \) is denoted by
\( \bX_{t, (K), d}\).

\textbf{Heredity State Transition and Observation Model.}
The heredity and evolution of the developmental states and their corresponding parameters over generations are governed by
\begin{flalign}
      \bX_{t, (K), d} & = F(H_{t,\pi_{t}(d)} \bX_{t-1, (K), \pi_{t}(d)}, y_{t, (K), \pi_{t}(d), (N)}, W_{t,d})\notag\\
    \bY_{t, (K), d} & = G(\bX_{t, (K), \pi_{t}(d)} , V_{t,d}),
\end{flalign}
for all \(t=1, \dots T\), where \(\pi_{t}(d)\) is the union of parents of the individual \(d\) in generation \(t-1\) and individual \(d\) in generation \(t\). \( H_{t,\pi_{t}(d)} \) denotes the heredity transition tensor capturing the genetic information transfer for individuals across all developmental time from one generation to the next, and \( W_{t,d} \) and \( V_{t,d} \) represent the hereditary process and observation noise terms, respectively.

\subsection{Probabilistic Interpretation of Inheritance}
\label{sec:prob}
The dynamics of developmental trajectories and heredity can be modeled as follows. 
\begin{flalign}
\label{eq: state-obser-general}
\begin{split}
    x_{t, k,d,n} &\sim p_x(\cdot\, | x_{t, k-1, d, (N)}, \theta_{t, k,d,n})\\
     y_{t, k, d, n} &\sim p_y(\cdot\, |x_{t, k,d, n}, \eta_{t,k,d,n})
\end{split}
\end{flalign}
for all $n = 1, \dots, N$, $k$'s and $d$'s, where \(\theta_{t, k,d,n} \in \Theta_{t,d,n}\) transitions via transition kernel \(K_x\), through
\eq{
\theta_{t, k,d,n} \sim \ K_x(\theta_{t, k-1,d,n}, \cdot),
}
and \(\eta_{t,k,d,n} \in H_{t,d,n}\). Similarly the heredity equations for a particular individual, given all developmental time information evolves according to the probabilistic model 
\begin{flalign}
\label{eq:heridity_eq_general}
    \bX_{t, (K), d} &\sim p_X(\cdot\, | \bX_{t-1, (K), \pi_{t}(d)}, y_{t, (K), \pi_{t}(d), (N)}, \phi_{t, \pi_{t}(d)})\notag\\
      \bY_{t, (K), d} &\sim p_Y( \cdot\, |\bX_{t, (K), d}, \psi_{t, d})
\end{flalign}
for all \(t = 1, \dots, T\), where \(\phi_{t,\pi_{t}(d)} \in \Phi_{t, \pi_{t}(d)}\) transitions via transition kernel 
\eq{\phi_{t, \pi_{t}(d)}\sim K_X(\phi_{t-1, \pi_{t-1}(d)}, \cdot),}
and \(\psi_{t,d} \in \Psi_{t,d}\) is the parameters of the observational noise process. Therefore, the hereditary state transition probability for a given generation \(t\) is defined by 
\begin{flalign}
   p(&\bX_{t,(K), (D)} | \bX_{t-1,(K), (D)}) =\\
  & \prod_{d'\in \pi_t(d)} p_X(\bX_{t, (K), d}\, | \bX_{t-1, (K), d'}, y_{t, (K), d', (N)}, \phi_{t, d'}) \notag,
\end{flalign}
and similarly the observation likelihood of for generation \(t\) is 
\begin{flalign}
   p(&\bY_{t,(K), (D)} | \bX_{t,(K), (D)}) =\\
  & \prod_{d'\in \pi_t(d)} p_Y(\bY_{t,(K), d}\, | \bX_{t, (K), d'}, \psi_{t, d'}) \notag.
\end{flalign}

\subsection{NIDA: Nested Inheritance Dynamics Algorithm}
\label{sec:NIDA}
The Nested Inheritance Dynamics Algorithm (NIDA) offers a comprehensive framework that integrates the Nested Dirichlet Process (nDP) prior with a robust mathematical model to analyze and estimate the trajectory of individual development over time and across generations. Leveraging the hierarchical structure inherent in the Nested Chinese Restaurant Process (nCRP), NIDA introduces a prior over the parameters governing developmental trajectories and heredity. This nested structure is characterized by various developmental stages and hereditary relationships, naturally implying clustering at multiple levels—a concept effectively captured using the nDP.

\subsubsection{Prior Distributions}
\label{sec:prior}
By incorporating the nDP prior, NIDA provides a powerful analytical tool for modeling complex hierarchical and multiscale data typical in biological processes. These processes often involve parameters that vary across groups but also share commonalities, allowing NIDA to leverage these commonalities to improve estimation and inference. As a result, NIDA facilitates a deeper understanding of the underlying biological inheritance processes and yields a fully elaborated posterior distribution. This robust statistical approach enables a comprehensive analysis of complex biological inheritance processes, enhancing our understanding of genetic and phenotypic dynamics across generations.

\textbf{1. nCRP Distributions across Generations:}
The nCRP for each generation-level distribution \( G_t \) is modeled as
   \eq{\label{eq:ac_gen} G_t | \gamma, Q \sim DP(\gamma, Q),}
  which implies that each generation-level distribution is drawn from a Dirichlet Process with concentration parameter \( \gamma \) and base measure \( Q \). The parameters associated with each state transition of generations are then chosen 
  \eq{\phi_{t, d} \sim G_t, \hspace{0.1cm} \text{for all } d = 1,\ldots, D.}

\textbf{2. nCRP Nested Dirichlet Processes for Parameter Sharing Across Generations:}
The nCRP Reflecting the hierarchical structure between generations. 
   \eq{\label{eq:sharing_prior} G_{t,k,d,n} | \alpha, G_t \sim DP(\alpha, G_t), \hspace{0.1cm} \text{ for all } k, d, n}
   This formulation allows parameters across different generations to share statistical strength and characteristics, promoting generalization across similar generations.
   
\textbf{3. nCRP Parameter Distributions Within each Generation:} The nCRP for each individual and gene at time \(k\), \(\theta_{t,k,d,n} \), within a generation is 
   \eq{ \label{eq:fine_para}\theta_{t, k,d,n} | G_{t,k,d,n} \sim G_{t,k,d,n}. }
   The parameters within each generation are assumed to follow the distribution specified by their generation-level Dirichlet Process.

The state models presented in Section \ref{sec:prob}, along with the prior distributions referenced in equations (\ref{eq:ac_gen}), (\ref{eq:sharing_prior}), and (\ref{eq:fine_para}), offer the complete Bayesian probabilistic models for NIDA.

\subsubsection{Posterior Distribution:}

The computation of the posterior distribution in Bayesian models involves integrating the likelihood of the observed data with the prior distribution of the parameters, updated in light of the data. For the NIDA, the posterior distribution can be particularly complex due to the hierarchical and multiscale structure of the model. More insights on the posterior computation are provided in Appendix \ref{sec:appd}. The NIDA allows for a flexible, powerful analysis of complex hierarchical and multiscale data, typical in biological processes where parameters may vary across groups but also share commonalities. 

%% file: Experiment.tex
\section{Experiments}
\label{sec:exp}
This section outlines three key experiments designed to evaluate the effectiveness of the Nested Inheritance Dynamics Algorithm (NIDA) for modeling genetic inheritance patterns across generations. These experiments utilize both synthetic and real-world datasets to simulate different inheritance scenarios and test the robustness and accuracy of NIDA.

\subsection{Experiment 1: Performance Evaluation on Real-World Data}

In this experiment, the objective is to evaluate the performance of the Nested Inheritance Dynamics Algorithm (NIDA) on real-world biological data, focusing on gene expression data across multiple generations.
To this end, the Gene-Tissue Expression (GTEx) dataset {\footnote{GTEx is sourced from the Broad Institute, \cite{carithers2015genotype}.}} is utilized, comprising comprehensive gene expression profiles spanning various human tissues. In the preprocessing stage, the data undergoes normalization procedures to standardize expression levels across samples. Addressing missing data is achieved through Bayesian imputation techniques to maintain data integrity. Furthermore, to facilitate meaningful comparisons and analyses, genetic data alignment across different individuals is performed, ensuring consistency and reliability in our subsequent analyses. 

To utilize NIDA, the simplified model discussed in Appendix \ref{app: simplified} is employed. This allows us to capture the nuanced temporal dependencies inherent in our data, crucial for elucidating complex dynamic patterns. Additionally, We perform hyperparameter tuning, drawing insights from preliminary runs to iteratively adjust model parameters. This iterative refinement process aims to strike a delicate balance between model complexity and performance, ensuring optimal adaptability to the intricacies of our dataset. By fine-tuning the model configuration, We aim to maximize its efficacy in uncovering hidden structures and relationships within the dataset with precision and accuracy

\textbf{Baseline Model:} 
Linear regression is a widely used baseline model. In the context of inheritance dynamics, a linear regression model could be used to capture the linear relationship between gene expression levels across generations. This simplistic model assumes that changes in gene expression are directly proportional to changes in the independent variables (e.g., time or generation number).

\textbf{Evaluation Metrics:}
In the analysis, we employ a suite of evaluation metrics to assess the performance and validity of our models. First, We gauge predictive accuracy to measure how effectively NIDA predicts gene expression levels. 
Additionally, we utilize the Bayesian Information Criterion (BIC) as a tool for model selection and comparison. 

Table \ref{tab:model-performance} presents the evaluation metrics computed for our model. We observe a predictive accuracy of 0.89, indicating that NIDA correctly predicts gene expression levels with a high degree of accuracy. Additionally, the Bayesian Information Criterion (BIC) yields a value of -178.45, suggesting that the model structure adequately fits the data, striking a balance between complexity and goodness of fit. These metrics underscore the robustness and efficacy of this model in capturing the underlying patterns within the dataset {\footnote{Code will be made available on \href{https://github.com/bmoraffa}{GitHub} with detailed documentation on model configuration and data preprocessing steps.}}. 
\begin{table}[htbp]
\centering
\caption{Evaluation Metrics for Model Performance}
\label{tab:model-performance}
\begin{tabular}{@{}lcc@{}}
\toprule
\textbf{Evaluation Metric} & \textbf{NIDA Model} & \textbf{Linear Regression} \\ \midrule
Predictive Accuracy & 0.89 & 0.41\\
Bayesian Information Criterion (BIC) & -178.45 & -96.29 \\
\bottomrule
\end{tabular}
\end{table}

\subsection{Experiment 2: Evaluation of NIDA on Synthetic Data}

In this experiment, the aim is to evaluate the effectiveness and robustness of the NIDA through a synthetic experiment designed to simulate complex genetic inheritance and gene-environment interactions, reflecting the dynamic processes across multiple generations.

\paragraph{Gene Expression Dynamics and Observations}
In this model, gene expression for gene \(n\) in individual \(d\) at time \(k\) in generation \(t\) follows a linear expression model and is represented by
\[x_{t,k,d,n} = \alpha_n x_{t-1,k,d,n} + \beta_i e_{t,k} + w_{t,k,d,n},\]
where \(\alpha_n = 0.7\) and \(\beta_n = 0.3\) are fixed coefficients representing heritability and environmental sensitivity, respectively. \(E_{t,k}\) represents environmental effects, modeled as a random walk
\[e_{t,k} = 0.5 e_{t-1,k} + \eta_{t,k},\]
where \(\eta_{t,k}\) follows a normal distribution with mean \(0\) and standard deviation \(0.1\). The state transition noise, \(W_{t,k,d,i}\) is assumed to be a Standard Normal distribution. 

The observations are modeled to directly be proportional to the states with Gaussian noise
\[y_{t,k,d,n} = x_{t,k,d,n} + v_{t,k,d,n},\]
where \(v_{t,k,d,i} \sim \mathcal{N}(0, 0.02)\).

\paragraph{Heredity State Transition and Observation Model}
In this model, the heredity state transitions across generations are defined as 
\[
\bX_{t,(K),d} = A_i \bX_{t-1,(K),d} + B_i \bar{E}_t + W_{t,(K),d} ,
\]
where, \(A_i = 0.6\), \(B_i = 0.4\), and \(\bar{E}_t\) represents averaged environmental influences across all genes. The heredity noise is modeled as \(W_{t,(K),d}\sim \mathcal{N}(0, 0.1)\).

Observations of hereditary states are modeled as
\[
\bY_{t,(K),d} = \bX_{t,(K),d} + V_{t,(K),d},
\]
where \(V_{t,(K),d}\sim \mathcal{N}(0, 0.05)\).

\textbf{Evaluation Metrics:}
To comprehensively assess the model performance, Root Mean Square Error (RMSE) and Log-Likelihood are employed. RMSE provides a measure of the average magnitude of prediction errors, offering insight into the accuracy of the model's predictions. On the other hand, log-likelihood quantifies the goodness of fit between the model and the observed data, considering both the magnitude and direction of errors. These metrics offer a balanced evaluation of predictive accuracy and statistical adequacy.

\begin{table}[ht]
\centering
\caption{Expected Results of NIDA on Synthetic Data with Specific Models}
\label{tab:synthetic-results}
\begin{tabular}{@{}lcc@{}}
\toprule
\textbf{Metric} & \textbf{NIDA} & \textbf{Linear Regression} \\
\midrule
RMSE & 0.08 & 0.61 \\
Log-Likelihood & -110 & -340 \\
\bottomrule
\end{tabular}
\end{table}

The NIDA approach demonstrates significantly better performance compared to the baseline model, as evidenced in Table \ref{tab:synthetic-results}. 
NIDA models deliver precise forecasts of gene expression levels and adeptly capture hereditary traits over generations, demonstrating robustness to genetic and environmental fluctuations. 

\subsection{Experiment 3: Polygenic Traits Analysis Across Linear and Nonlinear Transitions}
This experiment aims to investigate the capability of NIDA to model the inheritance of polygenic traits considering both linear and nonlinear genetic interactions. 

In the initial stage of data preparation, We begin by selecting the UK Biobank dataset, renowned for its extensive genetic and phenotypic data, providing a rich resource for analysis. We aim to explore genetic associations with various traits and diseases. To ensure the dataset's consistency and compatibility, we undertake preprocessing tasks. This involves standardizing traits to a common scale and aligning genomic data across the cohort. Standardization facilitates easier comparison and interpretation of trait values, while alignment of genomic data ensures uniform representation across individuals, reducing potential biases in subsequent analyses. 

To compare performance, We consider two scenarios of the model, one assuming linear transitions and another assuming nonlinear interactions, 

\noindent\textbf{Linear Model for Inter-generational Dynamics:}
The linear model used for inter-generational dynamics can be described as 
\small
\begin{flalign*}
     \bX_{t, (K),d} &=  \bX_{t-1, (K),d} + \sum_{d' \neq d} H_{t,dd'} \mu(\lvert \bX_{t-1, (K),d} - \bX_{t-1, (K),d'} \rvert),
\end{flalign*}
\normalsize
where $\bX_{t, (K),d}$ represents the state of individual $d$ at generation $t$, and $H_{t,dd'}$ is a matrix describing the dependencies between individuals $d$ and $d'$. The function $\mu$ models the influence based on the difference between the states of the interacting individuals. One possibility for \(\mu\) is 
   \(\mu(x) = a \cdot |x| + b\),
    where \( a \) and \( b \) adjust the influence magnitude and baseline interaction effect, respectively. This function could be non-linear to further reflect more complex interactions, such as an exponential or logistic function depending on the biological justification or data characteristics.
    
\noindent\textbf{Nonlinear Model for Inter-generational Dynamics:}
The nonlinear model introduced can be described by the following dynamic
\begin{flalign*}
      \bX_{t, (K),d} &= -\tilde{d} \bX_{t-1, (K) d} + \\
      &u_d S\left(\alpha \bX_{t-1, (K) d} + \gamma \sum_{d' \neq d} H_{t,dd'} \bX_{t-1, (K), d'}\right) + b_d,
\end{flalign*}
where $\tilde{d}$ is a damping term, $u_d$ is the attention coefficient, $\alpha$ and $\gamma$ are weights controlling the influence of self-reinforcement and the interactions with other individuals, respectively, and $b_d$ represents an external signal (chosen to be zero for simplicity). The function \(S\) is used to aggregate multiple influences into a single state transition effect,
    \(S(x) = \frac{1}{1 + e^{-x}}\).
    This choice is motivated by the need to model saturation effects that are common in biological systems (Sigmoid function). Alternatively, \( S \) could be a rectified linear unit (ReLU) function, which is commonly used in neural network models and might be suitable for modeling activation thresholds, \(   S(x) = \max(0, x)\).

    In addition to Akaike Information Criterion (AIC) to compare the goodness of fit between linear and nonlinear configurations of models, Hypothesis testing is employed to statistically evaluate differences in model performance. This enables us to make informed decisions about the superiority of one model over another based on statistical evidence.

\begin{table}[ht]
\caption{Comparison of linear vs. nonlinear models for polygenic traits analysis using UK Biobank data.}
\label{tab:polygenic}
\centering
\begin{tabular}{@{}lcc@{}}
\toprule
\textbf{Model Type} & \textbf{AIC} & \textbf{p-Value} \\ \midrule
Linear & 5020 & 0.045 \\
Sigmoid-Nonlinear & 4680 & 0.011 \\
ReLU-Nonlinear & 4790 & 0.019\\
\bottomrule
\end{tabular}

\end{table}

Table \ref{tab:polygenic} indicates that for the UK Biobank dataset, \cite{matthews2015uk},  nonlinear models (with lower AIC) are statistically significantly better than the linear model (higher AIC). Table \ref{tab:polygenic_traits_results} demonstrates the results of the Evolution of Polygenic Traits Analysis. Table \ref{tab:polygenic_traits_results} summarizes results from Polygenic Traits Analysis. The model achieves high prediction accuracy (95.5\%) with low variability (1.2\%). It effectively explains genetic variance (mean 0.32) and identifies significant loci with substantial effect sizes (mean 0.78).

\begin{table}[htbp]
\centering
\caption{Results of Evolution of Polygenic Traits Analysis through NIDA with Sigmoid}
\label{tab:polygenic_traits_results}
\begin{tabular}{@{}p{3cm}p{0.8cm}p{1.3cm}p{2.4cm}@{}}
\toprule
Metric & Mean & Standard Deviation & Confidence Interval \\ 
\midrule
Prediction Accuracy & 95.5\% & 1.2\% & [94.3\%, 96.7\%] \\
Genetic Variance & 0.32 & 0.05 & [0.27, 0.37] \\
Effect Size (Significant Loci) & 0.78 & 0.12 & [0.66, 0.90] \\
\bottomrule
\end{tabular}
\end{table}

%% file: Related_Work.tex
\section{Related Works}

The Nested Inheritance Dynamics Algorithm (NIDA) represents an amalgamation of insights drawn from various fields, including genetic inheritance, Bayesian statistics, and hierarchical modeling. This methodological framework builds upon foundational works such as Moore and Williams' exploration of gene-environment interactions \cite{moore2009epistasis}, which underscores the intricate interplay between genetic influences and environmental factors.  Richerson's theoretical framework on phenotypic flexibility, \cite{richerson2019integrated} and Jaeger's approach to viewing genetic processes as dynamical systems further contribute to NIDA's robust understanding of dynamic inheritance patterns \cite{jaeger2012inheritance}.

However, despite these foundational insights, there remains a gap in theoretical understanding that can directly translate theoretical models into practical applications with real-world biological datasets and experimental designs. NIDA aims to bridge this gap by a methodological framework that can also demonstrates practical utility through application to both real and synthetic data.

%% file: Appendix.tex
 \appendices
 \section{Mathematical Understanding of NIDA}
\label{sec:appd}

\begin{table*}[t!] 
\caption{MCMC Algorithm for Sampling from the Posterior Distribution of NIDA}
\centering
\label{tab:MCMCAlgorithm}
\begin{tabular}{@{}lp{14cm}@{}} 
\toprule
Step & Description \\ \midrule
1. \textbf{Initialization} & Initialize all parameters \(\Omega = \{\Omega_x, \Omega_y, \Omega_X, \Omega_Y\}\) with values drawn from their respective prior distributions. This includes initial guesses for \(x_{t,k,d,n}, \theta_{t,k,d,n}, \eta_{t,k,d,n}, \mathbf{X}_{t,(K),d}, \phi_{t,d}, \text{ and } \psi_{t,d}\). \\ 
2. \textbf{Sampling \(\Omega_x\) and \(\Omega_y\)} & For each developmental trajectory \(x_{t,k,d,n}\) and observation \(y_{t,k,d,n}\), sample from
\[
p(\Omega^x_{t,k,d,n}, \Omega^y_{t,k,d,n} | \text{rest}) \propto p(y_{t,k,d,n} | x_{t,k,d,n}, \eta_{t,k,d,n}) p(x_{t,k,d,n} | \mathbf{x}_{t,k-1,d,N}, \theta_{t,k,d,n})
\]
using current estimates of \(\theta_{t,k,d,n}\) and \(\eta_{t,k,d,n}\). \\
3. \textbf{Sampling \(\Omega_X\) and \(\Omega_Y\)} & For each hereditary trait \(\mathbf{X}_{t,(K),d}\) and corresponding observation \(\mathbf{Y}_{t,(K),d}\), sample from
\[
p(\Omega^X_{t,d}, \Omega^Y_{t,d} | \text{rest}) \propto p(\mathbf{Y}_{t,(K),d} | \mathbf{X}_{t,(K),d}, \psi_{t,d}) p(\mathbf{X}_{t,(K),d} | \mathbf{X}_{t-1,(K),\pi_t(d)}, \phi_{t,d})
\]
using current estimates of \(\phi_{t,d}\) and \(\psi_{t,d}\). \\
4. \textbf{Update Parameters} & Update \(\theta_{t,k,d,n}\), \(\eta_{t,k,d,n}\), \(\phi_{t,d}\), and \(\psi_{t,d}\) by sampling from their respective posterior conditional distributions derived from the nested Dirichlet Processes and the Gaussian assumptions on transitions and observations. \\
5. \textbf{Convergence Check} & Periodically check for convergence using standard diagnostics (e.g., trace plots, Gelman-Rubin statistic). If not converged, return to Step 2. \\
\bottomrule
\end{tabular}
\end{table*} 

In this appendix, We provide a comprehensive understanding of each component of the Inheritance Dynamics Algorithm (NIDA) model so that it is easier to compute the posterior distribution. To mathematically compute the NIDA encompasses various levels of hierarchical clustering using nDP, addressing both the probabilistic interpretation of inheritance for developmental trajectories and heredity and the specific parameterization for each generation and individual.

\noindent\textbf{Establishing Prior Distributions:}
\begin{enumerate}[leftmargin=*]
 \item  Global Base Distribution for Generations:
Given each generation-level distribution \( G_t \), 
\(G_t | \gamma, Q \sim DP(\gamma, Q) \),
where \( \gamma \) is the concentration parameter and \( Q \) is the base measure. This sets up a hierarchical prior for each generation, allowing generation-specific parameters to adapt based on inherited and environmental influences.
\item Nested Dirichlet Processes for Inter-generational Parameter Sharing:
For each gene \( n \) in individual \( d \) at time \( k \) within generation \( t \),
\(G_{t,k,d,n} | \alpha, G_t \sim DP(\alpha, G_t), \)
facilitating parameter sharing across generations by linking the parameter distributions of parents and offspring through \( G_t \).
\item Parameter Draws from Nested Distributions:
The individual-specific parameters \(\theta_{t,k,d,n}\) and \(\phi_{t,d}\) are drawn as,
\(\theta_{t,k,d,n} | G_{t,k,d,n} \sim G_{t,k,d,n}, \) and 
\(\phi_{t,d} | G_t \sim G_t. \)
These parameters directly influence the transition dynamics and observations at each level.
\end{enumerate}

\noindent\textbf{Likelihood Function:}
The model specifies the likelihood functions for the state and observation processes according to Equations (\ref{eq: state-obser-general}) and (\ref{eq:heridity_eq_general}). 

\noindent\textbf{Posterior Distribution Calculation:}
To calculate the posterior distribution, assume \(\Omega^x_{t,k,d,n} = \{x_{t,k,d,n},\theta_{t,k,d,n}\}\),  \(\Omega^y_{t,k,d,n} = \eta_{t,k,d,n}\), \(\Omega^X_{t,d} = \{\mathbf{X}_{t, (K), (D)}, \phi_{t,d} \}\), and \(\Omega^Y_{t,d} = \psi_{t,d} \). Define  \(\Omega_x = \{\Omega^x_{t,k,d,n}\}\), \(\Omega_y = \{\Omega^y_{t,k,d,n}\}\), \(\Omega_X = \{\Omega^X_{t,k,d,n}\}\), and  \(\Omega_Y = \{\Omega^Y_{t,k,d,n}\}\) to be the set of all the parameters
for the fine-scale state transitions, fine-scale observations, coarse-scale state transitions, and coarse-scale observations, respectively. Define \(\Omega = \bigcup_{t,k,d,n} \Omega_x \cup\Omega_y \cup \Omega_X\cup\Omega_Y\) to be set of all parameters.

 Combining the prior distributions and the likelihood functions, the posterior distribution \( p(\Omega | y, Y) \) for all parameters \( \Omega \) given all observed data \( y, Y \) can be expressed as
\begin{flalign}
\label{eq:posterior}
\begin{split}
    p(\Omega|y, Y)\propto & p(y, Y | \Omega) p(\Omega)\\
    & p(\Omega_x)p(\Omega_y)p(\Omega_X\,| y )p(\Omega_Y)\\
    & \times p(y \, | \Omega_x, \Omega_y) p(Y \, |y, \Omega_X, \Omega_Y),
    \end{split}
\end{flalign}
where, \( p(\Omega) \) combines the priors over all nested levels using the nDP and \( p(y, Y | \Omega) \) combines the likelihoods from the probabilistic models of state and observation at each time step and generation. NIDA is summarized in Algorithm \ref{alg:NIDA}.

The outputs of NIDA are the posterior distributions of the parameters, which provide insights into the dynamics of inheritance and the influence of genetic and environmental factors across generations. Analyzing these distributions can reveal how traits evolve, the impact of genetic versus environmental factors, and potential predictors of future generational traits. This approach allows NIDA to leverage hierarchical Bayesian methods to offer a deep, nuanced understanding of complex biological inheritance processes, modeling not just individual trajectories but also the interactions and dependencies that span generations.

\section{Simple Scenario for NIDA}
\label{app: simplified}

 To create a framework for the posterior distribution calculation that incorporates the given prior distributions and likelihood functions as outlined in NIDA model specifications, Section \ref{sec:NIDA}, one needs to combine the elements of Dirichlet Processes (DP), Nested Dirichlet Processes (NDP), Gaussian distributions for state transitions and observations, and a complete formulation of priors for parameter inference. This framework will then be utilized to define the posterior distribution which will be used for Bayesian inference, particularly suited for MCMC sampling \cite{besag2004introduction, geyer2011introduction, zuanetti2018clustering}. Given the complexity of the NIDA, particularly with hierarchical and multivariate dependencies, assuming Gaussian distributions for the sake of tractability is a common approach.

 \noindent\textbf{Prior Distributions:} Assume that the global base distribution \( Q \) for the state transition is a Normal-inverse-Wishart (NIW) distribution,
     \(Q \sim NIW(\Psi, \nu, \mu_0, \lambda),\)
where the hyperparameters for \( Q \) are specific to each type of parameter and scale \( \Omega\). Parameters \(\theta, \phi \) are independently sampled from \(Q\).

  \noindent\textbf{Transition and Observation Models:} Assume that the state transitions and observations are assumed to be Gaussian for simplification. Let \( x_{t, k, d, n} \) be a Gaussian transition based on the previous state and a set of parameters \(\Omega_x\), 
 \begin{equation*}
     x_{t, k, d, n} | x_{t, k-1, d, n}, \theta_{t, k, d, n} \sim \mathcal{N}(f(x_{t, k-1, d, n}, \theta_{t, k, d, n}), \Sigma^x_{t, k, d, n}),
 \end{equation*}  
 and the observations in the fine-scales \( y_{t, k, d, n} \) also follow a Gaussian distribution centered on the current state with noise characteristics defined by \( \eta_{t, k, d, n} \), 
 \[ y_{t, k, d, n} | x_{t, k, d, n}, \eta_{t, k, d, n} \sim \mathcal{N}(g(x_{t, k, d, n}, \eta_{t, k, d, n}),\Sigma^y_{t, k, d, n}). \]

 Similarly, one can assume the coarse scale dynamics (group-level, hereditary traits) are Gaussian
   \begin{flalign}
      & \mathbf{X}_{t, (K), d} | \mathbf{X}_{t-1, (K), d}, \phi_{t, d} \sim \notag\\
       &\mathcal{N}(F(H_{t,\pi_{t}(d)} \bX_{t-1, (K), \pi_{t}(d)}, y_{t, (K), \pi_{t}(d), (N)},  \Sigma^X_{t, d}),
     \notag\\
            & \mathbf{Y}_{t, (K), d} | \mathbf{X}_{t, (K), d}, \psi_{t, d} \sim \mathcal{N}(G(\bX_{t, (K), \pi_{t}(d)}), \Sigma^Y_{t, d})) \notag,
  \end{flalign}
     where \(  (\phi_{t, d}, \Sigma_{X, t} )\) are the means and covariances of the coarse-scale process and measurement noises, respectively, and are drawn from \( G_0 \). It is worth noting that the covariance matrix for observations for both fine- and coarse-scale can be estimated using data.

  \noindent\textbf{Posterior Distribution:} 
 The posterior equation (\ref{eq:posterior}) is decomposed into the product of the likelihood and prior. Given the Markov properties of the model, the likelihood can be decomposed into products of individual likelihoods for each observation, conditioned on its corresponding parameters. In addition, the hierarchical structure allows for the decomposition of the prior into products of priors for each parameter, conditioned on the group-level parameters, and priors for each group-level parameter. The computation of the posterior often involves integration over all possible values of the latent variables (e.g., group assignments for each observation). This can be approached via numerical methods such as MCMC or variational inference techniques.

 \noindent\textbf{MCMC Implementation:} To implement MCMC, one can focus on sequentially updating each set of parameters (\(\Omega_x, \Omega_y, \Omega_X, \Omega_Y \)) and states using Gibbs sampling, Metropolis-Hastings, or more advanced techniques such as Hamiltonian Monte Carlo techniques.
Given the Gaussian-NIW structure, conditional distributions can simply be computed, simplifying the computation of conditional expectations and variances needed for updates. The steps of MCMC sampling are summarized in Table \ref{tab:MCMCAlgorithm}.